\def\year{2022}\relax
\documentclass[letterpaper]{article} 
\usepackage{aaai22}  
\usepackage{times}  
\usepackage{helvet}  
\usepackage{courier}  
\usepackage[hyphens]{url}  
\usepackage{graphicx} 
\usepackage{natbib}  
\usepackage{caption} 
\DeclareCaptionStyle{ruled}{labelfont=normalfont,labelsep=colon,strut=off} 
\usepackage{algorithm}
\usepackage{algorithmicx}
\usepackage{algpseudocode} 

\usepackage{times}
\usepackage{epsfig}
\usepackage{graphicx}
\usepackage{amsmath}
\usepackage{amssymb}
\usepackage{multirow}
\usepackage{microtype}
\usepackage{graphicx}
\usepackage{subfigure}
\usepackage{booktabs} 
\usepackage{epsfig}
\usepackage{graphicx}
\usepackage{amsmath}
\usepackage{amssymb}

\usepackage{indentfirst}

\newcommand{\liuhao}{\fontsize{8.pt}{\topskip}\selectfont}
\newcommand{\moren}{\fontsize{9.5pt}{\baselineskip}\selectfont}

\usepackage{newfloat}
\usepackage{listings}
\floatstyle{ruled}
\newfloat{listing}{tb}{lst}{}
\floatname{listing}{Listing}
\title{Rethinking the Optimization of Average Precision: \\Only Penalizing Negative Instances before Positive Ones is Enough}
\author{
    Zhuo Li$^{1,2}$\thanks{Work done during an internship at Meituan.}, Weiqing Min$^{1,2}$\thanks{Corresponding author.}, Jiajun Song$^{1,2}$, Yaohui Zhu$^{1,2}$, Liping Kang$^{3}$,\\ Xiaoming Wei$^{3}$, Xiaolin Wei$^{3}$ and Shuqiang Jiang$^{1,2}$
}
\affiliations{
    $^1$Key Lab of Intell. Info. Process., Inst. of Comput. Tech., CAS, Beijing, China\\
    $^2$University of Chinese Academy of Sciences, Beijing, China\\
    $^3$Meituan, Beijing, China \\
    { \{zhuo.li, jiajun.song, yaohui.zhu\}@vipl.ict.ac.cn,}{ \{minweiqing,sqjiang\}@ict.ac.cn,}\\
    { \{kangliping, weixiaoming, weixiaolin02\}@meituan.com}
}

\begin{document}

\maketitle

\begin{abstract}
Optimizing the approximation of Average Precision (AP) has been widely studied for image retrieval. Limited by the definition of AP, such methods consider both negative and positive instances ranking before each positive instance. However, we claim that only penalizing negative instances before positive ones is enough, because the loss only comes from these negative instances. To this end, we propose a novel loss, namely Penalizing Negative instances before Positive ones (PNP),  which can directly minimize the number of negative instances before each positive one. In addition, AP-based methods adopt a fixed and sub-optimal gradient assignment strategy. Therefore, we systematically investigate different gradient assignment solutions via constructing derivative functions of the loss, resulting in PNP-I with increasing derivative functions and PNP-D with decreasing ones. PNP-I focuses more on the hard positive instances by assigning larger gradients to them and tries to make all relevant instances closer. In contrast,  PNP-D pays less attention to such instances and slowly corrects them. For most real-world data, one class usually contains several local clusters. PNP-I blindly gathers these clusters while PNP-D keeps them as they were. Therefore, PNP-D is more superior. Experiments on three standard retrieval datasets show consistent results with the above analysis. Extensive evaluations demonstrate that PNP-D achieves the state-of-the-art performance. Code is available at \url{https://github.com/interestingzhuo/PNPloss}
\end{abstract}
\section{Introduction}
\noindent Image retrieval \cite{Radenovic-TPAMI18,2020Two-stage} refers to finding all images containing relevant content with the query from the database. One important issue is to design the training objective. Recently, some works propose to directly optimize the Average Precision (AP) \cite{2019ap,2020smoothap,2019fastap} to achieve end-to-end training. Considering AP is non-differentiable and can’t be directly optimized by the gradient descent, these works leverage some transformations to approximate AP to make it differentiable. These methods greatly promote the development of retrieval. 

\label{intro}
\begin{figure}[t]
	\centering
	\includegraphics[width=7.5cm]{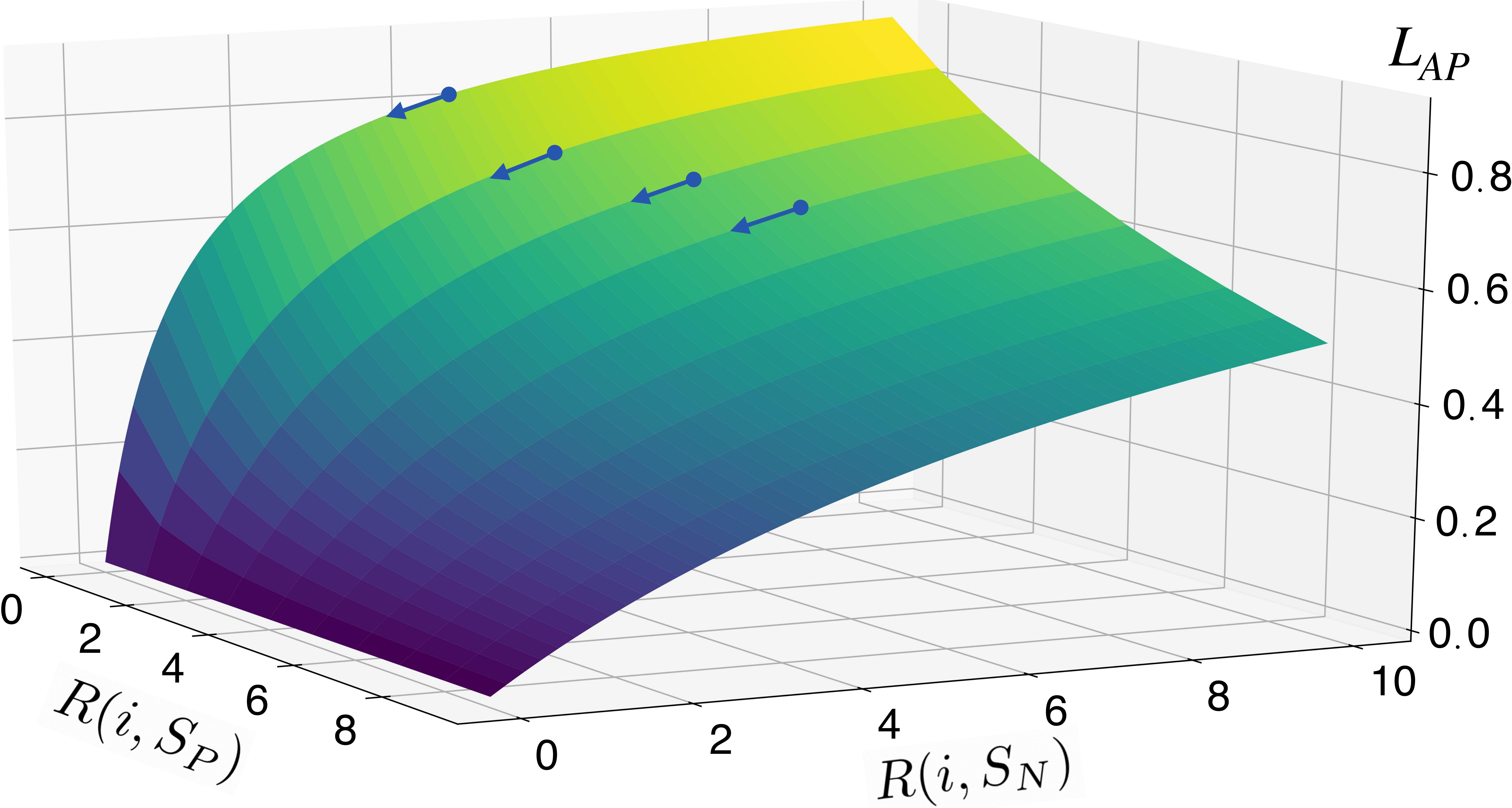}
	\caption{The functional image of AP-based loss $L_{AP}$: $R(i,S_P)$  and $R(i,S_N)$ are the number of positive and negative instances before the instance  $i$, respectively, where $S_P$ is the similarity set between the query and positive instances and $S_N$ is the similarity set between the query and negative instances. Arrows are the optimal path of positive instances. } 
	\label{AP}
\end{figure}

\begin{figure*}[htbp]

\centering

\includegraphics[width=15cm]{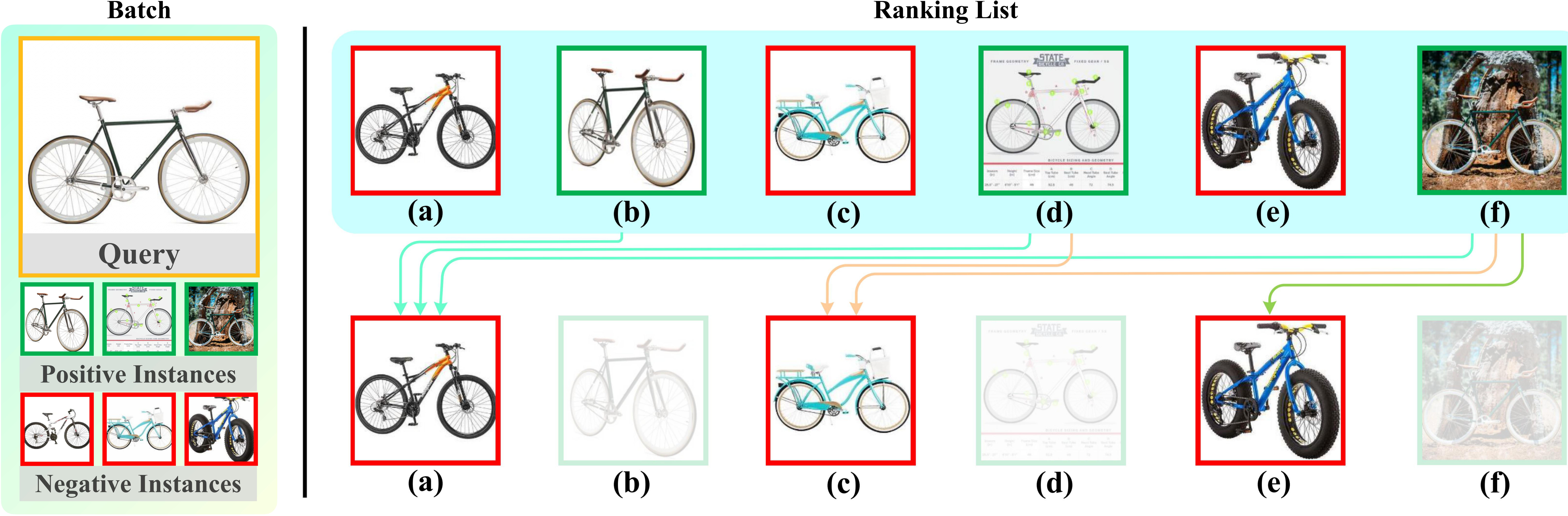}

\caption{Penalizing Negative instances before Positive ones (PNP): PNP directly improves the retrieval performance by penalizing negative instances before positive ones. Each arrow indicates the penalty of each negative instance. Each negative instance will receive penalties from all positive ones after it, thus the negative instance before more positive ones will receive stronger penalty and be quickly corrected.}

\label{fig:framework}

\end{figure*}

However, there are two problems for AP-based methods: {\textbf{(1) Redundancy exists in the optimization of AP.}} Fig.~\ref{AP} shows the functional image of AP-based loss $L_{AP}$. As shown in Fig.~\ref{AP}, the goal of minimizing $L_{AP}$ is equivalent to minimize $R(i,S_N)$, because $L_{AP} = 0$ only when $R(i,S_N)$ = 0. Therefore, the computation of $R(i,S_P)$ is redundant. \textbf{(2) Sub-optimal gradient assignment strategy.} Limited by the definition of AP, AP-based methods only adopt one specific gradient assignment strategy. Specifically, AP-based losses assign smaller gradient to larger $R(i,S_N)$, as shown in Fig.~\ref{AP}. Different gradient assignment strategies may result in different performance. Therefore, it leaves room for us to systematically explore different gradient assignment solutions to find potential better ones.



For the first problem, we propose a novel loss, namely Penalizing Negative instances before Positive ones (PNP), which directly minimizes the number of negative instances before each positive one. Therefore, each negative instance will receive penalties from all positive ones after it. In terms of each negative instance, if it ranks before more positive ones, it will receive more penalties and be corrected more quickly. As shown in Fig.~\ref{fig:framework}, image (a) will receive more penalty because there are more positive instances after it and it will be quickly corrected. In contrast, image (e) will be slowly corrected because that there are fewer positive instances (image (f)) after it.


For the second problem, we systematically investigate different gradient assignment solutions. We find that the derivative function of the loss function defines different gradient assignment solutions. To this end, we construct different derivative functions, which result in two different types of PNP, namely PNP-I with increasing derivative functions and PNP-D with decreasing ones.  PNP-I assigns larger gradients to the positive instances with more negative ones before and tries to gather all relevant samples together. In contrast, PNP-D assigns smaller gradients to such positive instances. Therefore, PNP-D will slowly correct these instances, which probably belong to another center of the corresponding category. As shown in Fig.~\ref{fig:framework}, PNP-I assigns larger gradients to image (f) and tries to quickly correct it, because there are more negative instances before it. In contrast, PNP-D assigns smaller gradient to image (f) by considering it probably belongs to another center of the corresponding category. For the real-world data, one class usually contains several local clusters, and PNP-D is more suitable for such case, thus may be more superior. Extensive evaluation on three standard retrieval datasets shows that PNP-D achieves the state-of-the-art performance. 

In summary, our major contributions are as follows:
\begin{itemize}

\item We propose a novel loss function named PNP, which can improve the approximation of AP by simply ignoring positive instances before the target positive one.

\item To our knowledge, we are the first research to design loss functions via the construction of their derivative functions. Based on the proposed method, we systematically investigate different gradient assignment solutions and obtain more appropriate loss functions.

\item Experimental results on three benchmark datasets demonstrate that the proposed method achieves the state-of-the-art performance.
\end{itemize}

\section{Related Work}

\noindent \textbf{Image Retrieval.} How to obtain a compact image descriptor for retrieval has been widely studied \cite{Radenovic-TPAMI18, 2020Two-stage,2020local-eccv,2020cao-eccv}. Previous works bend their efforts for generating descriptors through an aggregation of local features, such as Fisher vectors \cite{2010fisher} and VLAD \cite{2013VLAD, 2010VLAD}. More recently, CNNs have made impressive progress for image retrieval \cite{2018NetVLAD, 2014neuralcodes, Gordo-DIR-ECCV2016, 2016CNNimageretrieval, 2014similarity}, because of their powerful nonlinear fitting capabilities. Besides studying the design of neural network structure to generate discriminative descriptors, there are also many researches on how to design the training objective \cite{2019ap,2006contrastive,2008softrank,2016npair,2020Circle,2019softtriple,2021proxy}, which brings out another research topic, deep metric learning.

\noindent \textbf{Deep Metric Learning.} \noindent Deep metric learning has been widely studied because of its important role in many tasks \cite{2021multidistance,2021self},  such as image retrieval. How to design the objective for such task is one important topic. A wide variety of loss functions have been proposed in previous works. These loss functions \cite{Schroff-FaceNet-CVPR2015, Ji-CDIR-MM2017, 2019sodeep} minimize the intra-class distance while maximize the inter-class distance by considering the constructed pairs, tuples, etc. For example, the triplet loss \cite{Schroff-FaceNet-CVPR2015} directly optimizes the relative ranking of positive and negative instances given an anchor. However, there are too many pairs or tuples, and most of them are useless for training. To solve this problem, many methods have been proposed, such as pretraining for classification, combining multiple losses \cite{2020a-diva,2020b-sharing} and using complex sample mining strategies \cite{2017margin}.
\begin{figure*}[htbp]
	\centering
	\includegraphics[width=16cm]{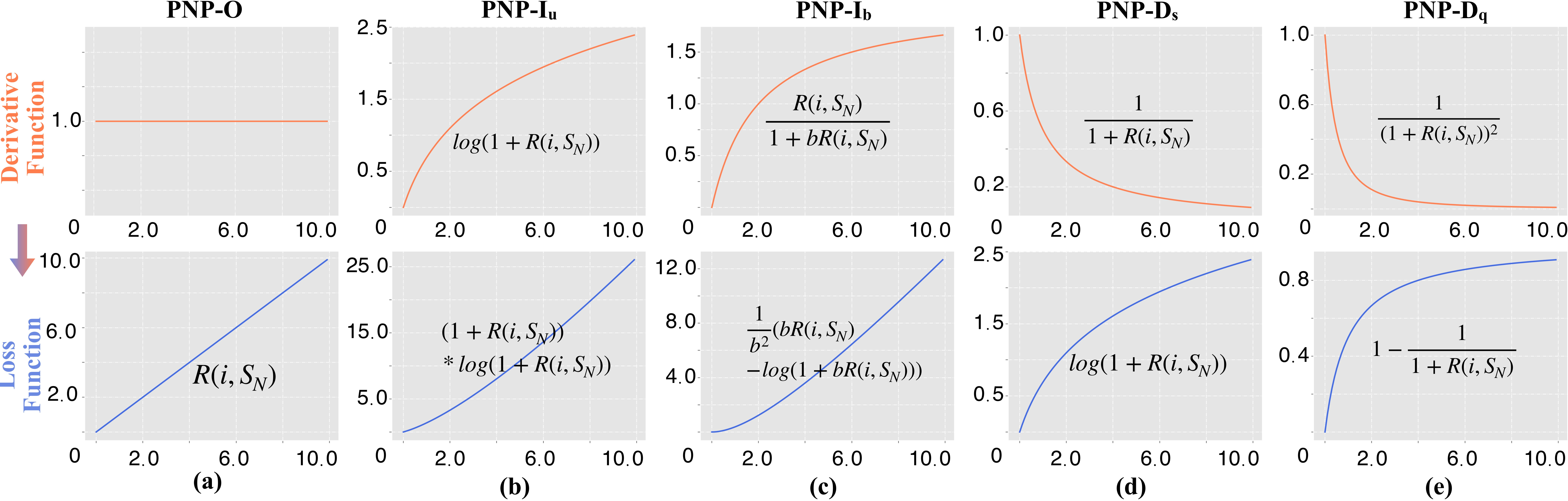}
	\caption{PNP Loss: Derivative functions and loss functions of different variants. PNP-I$_u$ and PNP-I$_b$ have increasing derivative functions, while PNP-D$_s$ and PNP-D$_q$ have decreasing derivative functions.}
	\label{fig:var}
\end{figure*}

\noindent \textbf{Optimizing Average Precision (AP).} Directly optimising AP has been widely studied in the retrieval community. Various methods \cite{2019fastap,2018hashing,2018localdescriptor,2019ap} have been proposed to overcome the non-differentiability in optimizing AP, such as optimizing a structured hinge-loss upper bound to the AP \cite{2018Efficient}, optimizing an approximation of AP derived from distance quantization \cite{2019fastap} and  smoothing the path towards optimization \cite{2020smoothap}. These approximations rely on the definition of AP to achieve better retrieval performance.

Different from above works, we propose the PNP loss, which directly minimizes the number of negative instances before each positive one to improve the retrieval performance. We further investigate different variants of PNP to explore potential better gradient assignment solutions via the construction of derivative functions. Extensive evaluations demonstrate that the proposed PNP loss achieves the state-of-the-art performance on three benchmark datasets.

\section{Methodology}
\noindent Given a query, the goal of retrieval systems is to rank all instances in a retrieval set $\Omega = \{I_i, i = 0, ..., m\}$ based on their similarities to the query, where $I_i$ is an instance and $m$ is the size of the retrieval set. For each query $I_q$, the set is split into a positive set $P_q$ and a negative set $N_q$, which are composed by instances of the same class and of different classes, respectively. 

For the query $I_q$, the similarity set $S_{\Omega}$ of all instances in the retrieval set are measured via a chosen similarity. In this paper, we use the cosine similarity, and $S_{\Omega}$ can be defined as:
\begin{equation}
	S_{\Omega} =\{ s_i =cos(v_q,v_i), i = 0, ... ,m \}
\end{equation}
where $v_q$, $v_i$  are the embeddings of $I_q$ and $I_i$, respectively. $S_{\Omega}$ can be divided into $S_P$ and $S_N$, \textit{i.e.}, $S_{\Omega} = S_P\cup S_N$, where $S_P=\{s_i, \forall i \in P_q\} $ and $S_N=\{s_i, \forall i \in N_q\} $ denote the positive and negative similarity set, respectively.

Based on the similarity set, the number of negative instances before $I_i$ can be defined as:
\begin{equation}
	R(i,S_N)=\sum_{j \in S_N, j \neq i} \mathbb I\{s_j - s_i > 0\}\label{R:function}
\end{equation}
where $\mathbb I \{\cdot\}$ is an Indicator function.

Note that $R(i,S_N)$ is not differentiable due to the Indicator function. Similar to \cite{2020smoothap}, we relax the Indicator function with a sigmoid function $G(\cdot;\tau)$, where $\tau$ refers to the temperature for adjusting the sharpness:
\begin{equation}
	G(x;\tau) = \frac{1}{1+e^{-\frac{x}{\tau}}}
	\label{g_function}
\end{equation}
Substituting $G(\cdot; \tau)$ into Eq.~\ref{R:function}, we can easily make $R(i,S_N)$ differentiable.

\subsection{Definition of PNP}
\label{PNP Loss}

\noindent AP-based loss is defined as Eq.~\ref{ap_f}. The goal of minimizing $L_{AP}$ is equivalent to minimize  $R(i,S_N)$, and thus the computation of $R(i,S_P)$ in Eq.~\ref{ap_f} is redundant. 
\begin{equation}
	L_{AP} = 1 - \frac{1}{|S_P|}\sum_{i \in S_P}\frac{1+R(i,S_P)}{(1+R(i,S_P)+R(i,S_N))}
	\label{ap_f}
\end{equation}
Therefore, we propose a PNP loss to directly minimize $R(i,S_N)$. An intuitive loss function $L_O$ can be defined as the number of negative instances before each positive one:

\begin{equation}
	L_O = \frac{1}{|S_P|}\sum_{i \in S_P}R(i,S_N)\label{pnp:basic}
\end{equation}
where $|S_P|$ is the normalization factor to remove the influence of the number of positive instances in each mini-batch. Substituting Eq.~\ref{R:function} into Eq.~\ref{pnp:basic}, we can get the original PNP. As shown in Fig.~\ref{fig:framework}, each negative instance will receive penalties from all the positive ones ranking after it. Therefore, negative instance with more positive ones after will receive more penalties and will be quickly corrected. The PNP-O and its derivative function are shown in Fig.~\ref{fig:var} (a), and the gradient assignment of PNP-O is constant.

\subsection{Variants of PNP}
\noindent Different gradient assignment strategies can result in different performance. In this section, we systematically investigate different gradient assignment solutions. Specifically, we design different variants of PNP by constructing their derivative functions, resulting in PNP-Increasing (PNP-I) with increasing derivative functions and PNP-Decreasing (PNP-D) with decreasing ones.

\subsubsection{PNP-Increasing (PNP-I)}

According to the above definition, we should guarantee that the derivative function of PNP-I is increasing.  Without losing generality, we can divide PNP-I into PNP-I$_u$ with unbounded derivative function and PNP-I$_b$ with bounded derivative functions.

As for PNP-I$_u$, we use Eq.~\ref{iop} as its derivative functions. The value of  this function will go down to zero when the loss is zero and go up to infinite when the loss tends to be infinity. Then we get PNP-I$_u$ via infinitesimal analysis, shown in Eq.~\ref{1_x:log_1_x}. $L_{I_u}$ has no hyper-parameters, and thus reduces the difficulty of training.

\begin{equation}
	\frac{\partial L_{I_u}}{\partial R(i,S_N)} = log(1+R(i,S_N))
	\label{iop}
\end{equation}
\begin{equation}
	L_{I_u} = \frac{1}{|S_P|}\sum_{i \in S_P}(1+R(i,S_N))*log(1+R(i,S_N))\label{1_x:log_1_x}
\end{equation}

As for PNP-I$_b$, in order to satisfy the two requirements: increasing and bounded, we use Eq.~\ref{b} as its derivative function. We introduce the parameter \textit{b} to achieve an adjustable boundary. Then we can also get PNP-I$_b$ as Eq.~\ref{1_b}.

\begin{equation}
	\frac{\partial L_{I_b}}{\partial R(i,S_N)} = \frac{R(i,S_N)}{1 +b R(i,S_N)}
	\label{b}
\end{equation}
\begin{equation}
	L_{I_b} = \frac{1}{|S_P|}\sum_{i \in S_P}\frac{1}{b^2}(bR(i,S_N) -log(1+bR(i,S_N)))\label{1_b}
\end{equation}

Loss functions and their derivative functions are shown in Fig.~\ref{fig:var} (b)\&(c). The derivative functions of these two functions are both increasing, which means that positive instances with more negative ones before will receive more penalties and these two loss functions will try to make such hard positive instances closer to the query.

\subsubsection{PNP-Decreasing (PNP-D)}

Specifically, we choose a common used decreasing function as the derivative functions of PNP-D, shown in Eq.~\ref{reci_pha} and Eq.~\ref{reci_pha1}. The introduced parameter $\alpha$ controls the speed of the descent, and larger $\alpha$ leads to faster speed. These two derivative functions result in PNP-D$_s$ with slow speed and PNP-D$_q$ with fast speed. After infinitesimal analysis, the PNP-D$_s$ and PNP-D$_q$ will be Eq.~\ref{log_1_x} and Eq.~\ref{reci}, respectively.

\begin{figure}[t]
	\centering
	\includegraphics[width=7cm]{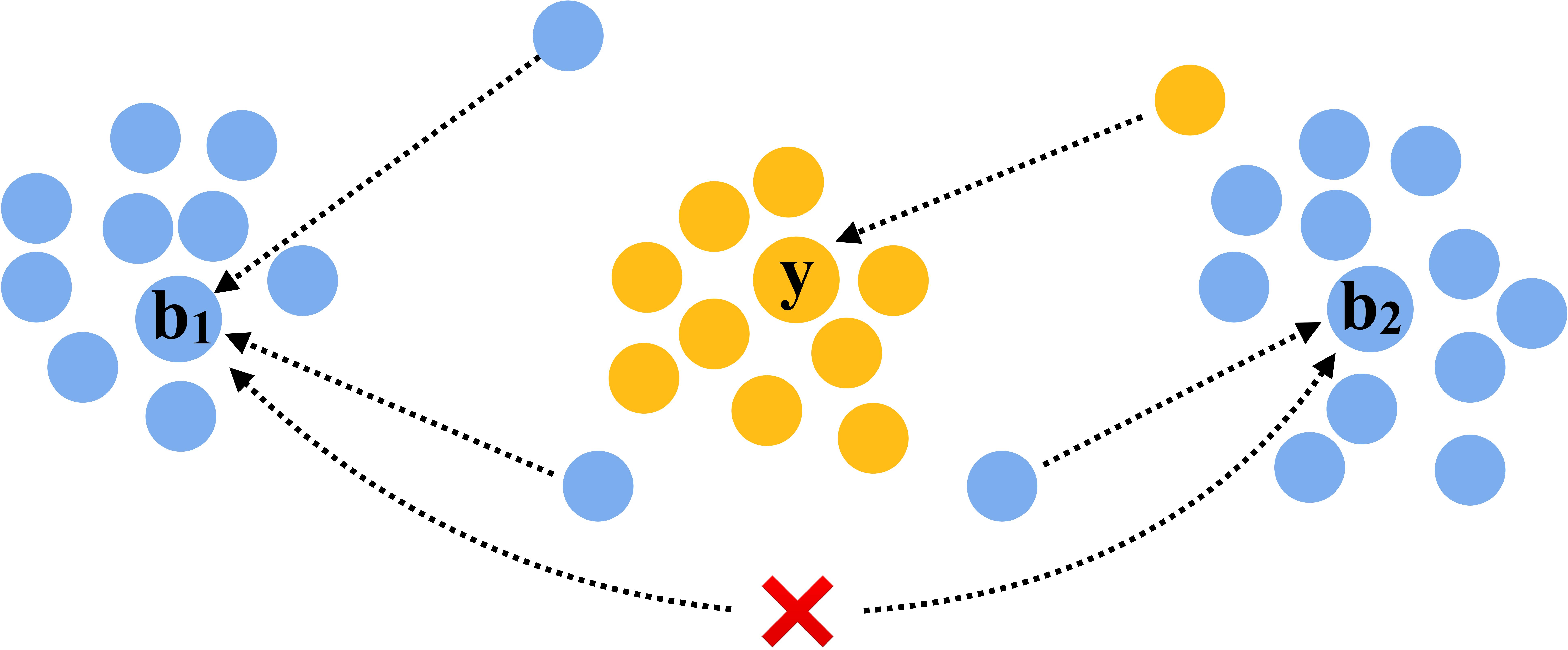}
	\caption{Comparison between PNP-I and PNP-D: The dashed arrows represent the directions of optimization. Different colors represent different categories. b and y represent centers of different categories. It is worth mentioning that b$_1$ and b$_2$ are different centers of class b. PNP-I blindly makes b$_1$ and b$_2$ closer while PNP-D  keeps such instances distribute into two clusters.}
	\label{toyexample}
\end{figure}

\begin{equation}
	\frac{\partial L_{D_s}}{\partial R(i,S_N)} = \frac{1}{(1+R(i,S_N))}
	\label{reci_pha}
\end{equation}
\begin{equation}
	\frac{\partial L_{D_q}}{\partial R(i,S_N)} = \frac{\alpha}{(1+R(i,S_N))^{\alpha+1}}\quad(\alpha \geq 1)
	\label{reci_pha1}
\end{equation}

\begin{equation}
	L_{D_s} = \frac{1}{|S_P|}\sum_{i \in S_P}log(1+R(i,S_N))
	\label{log_1_x}
\end{equation}
\begin{equation}
	L_{D_q} = 1-\frac{1}{|S_P|}\sum_{i \in S_P}\frac{1}{(1+R(i,S_N))^{\alpha}}\quad(\alpha \geq 1)
	\label{reci}
\end{equation}

Loss functions and their derivative functions of PNP-D are shown in Fig.~\ref{fig:var} (d)\&(e). The derivative functions of these two functions are both decreasing, which means that positive instances with fewer negative ones before will receive more penalties and these two loss functions will quickly correct such instances.

\begin{table*}[t]
	
	\liuhao
	
	\centering
	

	\setlength{\tabcolsep}{4.7mm}{
	\begin{tabular}{l|cccc|ccc}
		
		\bottomrule
		
		\multicolumn{1}{c|}{Method} & R@1 (\%) & R@10 (\%) & R@100 (\%) & R@1K (\%) & Dists@intra & Dists@inter & NMI (\%)\\ \hline
		
		PNP-O &77.9 &90.5 &96.3 &98.9 &0.318 &0.793 &90.0 \\ 
		
		PNP-I$_u$ &76.3&89.6&96.0&98.9&0.301&0.771&89.6\\ 
		
		PNP-I$_b$ &77.9&90.4&96.3&98.9&0.302&0.756&90.0\\ 
		
		PNP-D$_s$ &79.6&91.4&96.6&\textbf{99.0}&0.363&0.849&90.3\\

		PNP-D$_q$ &\textbf{80.1}&\textbf{91.5}&\textbf{96.7}&\textbf{99.0}&\textbf{0.380}&\textbf{0.946} &\textbf{90.4}\\ \bottomrule

	\end{tabular}}
	\caption{Results on different variants of the PNP loss on SOP\ (BS = 112). }
	\label{form_PNP}
	
\end{table*}
\begin{table*}[t]
	
	\liuhao
	
	\centering

	\setlength{\tabcolsep}{4.3mm}{
	\begin{tabular}{c|cc|cc|cc|ccc}
		
		\bottomrule
		
		\multirow{2}{*}{Method}
		
		&\multicolumn{2}{c}{Small (\%)} & \multicolumn{2}{c}{Medium (\%)} &\multicolumn{2}{c|}{Large (\%)}&\multirow{2}{*}{Dists@intra}&\multirow{2}{*}{Dists@inter}&\multirow{2}{*}{NMI (\%)} \\ \cline{2-7}
		
		& R@1& R@5 &R@1& R@5 &R@1& R@5& & & \\ \hline
		
		\multicolumn{1}{l|}{PNP-O} & 93.6 &97.1&91.6& 95.9&89.2&95.4&0.165&0.791&90.5\\ 
		
		\multicolumn{1}{l|}{PNP-I$_u$} & 92.4&96.9&90.4& 95.4&87.6&94.8&0.137&0.727&90.0\\ 
		
		\multicolumn{1}{l|}{PNP-I$_b$} &94.0&97.3&92.0&96.0&89.9&95.4&0.144&0.674&90.6 \\ 
		
		\multicolumn{1}{l|}{PNP-D$_s$} & 94.3& 97.3& 92.7& \textbf{96.4}& 91.0& 95.9& 0.203&0.865&90.8 \\
		
		\multicolumn{1}{l|}{PNP-D$_q$} &\textbf{94.6}&\textbf{97.4}&\textbf{92.9}&96.3&\textbf{91.4}&\textbf{95.9}&\textbf{0.204}&\textbf{0.974}&\textbf{91.0}\\ \bottomrule

	\end{tabular}}
	\caption{Results on different variants of the PNP loss on {VehicleID}\ (BS = 112).}
	\label{form_vehicle}
	
\end{table*}

\subsubsection{Discussion} 

PNP-I assigns larger gradients to positive instances with more negative ones  before, while PNP-D assigns smaller gradients to such positive instances. PNP-I tries to make all the relevant instances together. In contrast, PNP-D only quickly corrects positive instances with fewer negative ones before, because such samples are considered to belong to the same center with the query. As shown in Fig.~\ref{toyexample}, if $b_{1}$ and $b_{2}$ are sampled into one batch. PNP-I will try to make them gather by assigning larger gradients. In contrast, considering too many negative instances (yellow) between them, PNP-D will assign smaller gradients to them and keep the multi-center distribution for the blue class. PNP-D can satisfy such distribution which is also mentioned by \cite{2019softtriple}, and thus is superior to PNP-I. Such analysis has been further verified via our comprehensive experimental evaluation.

We just conduct above four variants of PNP. Actually, our proposed design strategy is general and  similar strategies can be further explored to find more solutions.

\begin{figure}[t]
	\centering
	\includegraphics[width=6.5cm]{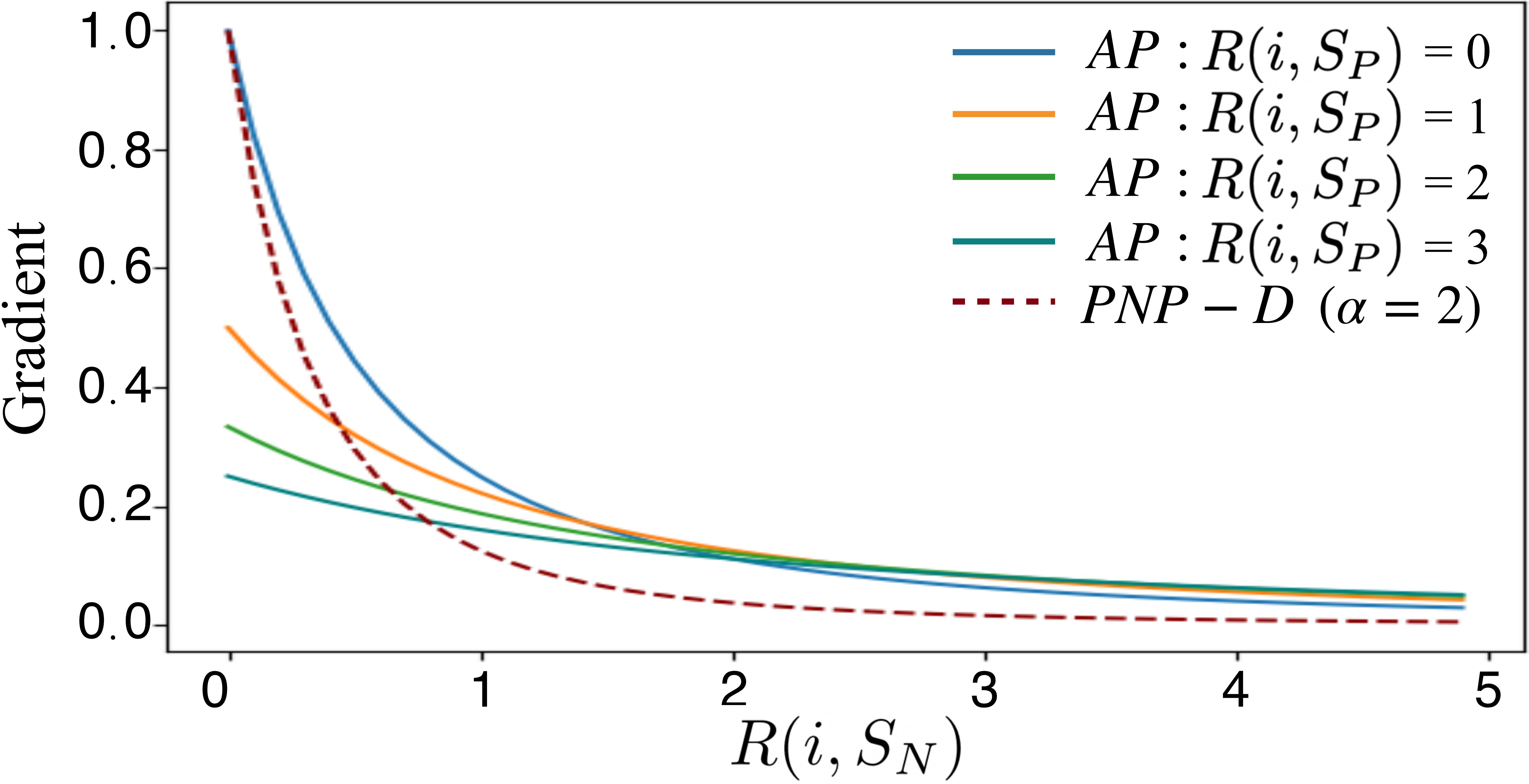}
	\caption{Relation between PNP and AP-based loss: The gradients of AP-based loss with respect to $R(i,S_N)$ when $R(i,S_P)$ = 0,1,2,3. The gradients of PNP-D ($\alpha$ = 2) is steeper than AP-based loss.} 
	\label{relationAP}
\end{figure}

\subsection{Relation between PNP and AP}

\label{dicussion}

\noindent In this section, we investigate the relation between PNP and AP. The derivative function of AP-based loss with respect to $R(i,S_N)$ is shown in Eq.~\ref{APn} and its functional image is shown in Fig.~\ref{relationAP}. AP-based methods have similar gradient assignment with PNP-D.
\begin{equation}
	\frac{\partial L_{AP}}{\partial R(i,S_N)} = \frac{1 + R(i,S_P)}{(1 + R(i,S_P) + R(i,S_N))^2}
	\label{APn}
\end{equation}
However, AP-based methods use different gradient assignments by considering the positions of positive instances. When too many positive instances rank before the target positive instance ($R(i,S_P) = 3$), the gradient respect to $R(i,S_N)$ is almost to be equal for different $R(i,S_N)$. It weakens the advantage of such gradient assignment, probably leading to worse retrieval performance for AP. In contrast, PNP can further enhance such advantage and finally achieve competitive performance.


\section{Experiment}

\subsection{Dataset}

\noindent We use the following three popular datasets as the benchmark datasets.

\noindent \textbf{Stanford Online Products (SOP) }\cite{Song-DML-CVPR2016} contains 120,053 product images divided into 22,634 classes. The training set contains 11,318 classes with 59,551 images and the rest 11,316 classes with 60,502 images are for testing.

\noindent \textbf{VehicleID }\cite{liu2016deep} contains 221,736 images of 26,267 vehicle categories, where  13,134 categories with  110,178 images are used for training. Following the same test protocol as \cite{liu2016deep}, three test sets of increasing sizes are used for evaluation (termed small, medium, large), which contain 800 classes (7,332 images), 1,600 classes (12,995 images) and 2,400 classes (20,038 images), respectively.

\noindent \textbf{INaturalist }\cite{van2018inaturalist} is a large-scale animal and plant species classification dataset with 461,939 images from 8,142 classes. We follow the  setting from \cite{2020smoothap} by keeping 5,690 classes for training, and 2,452 unseen classes for testing.

\subsection{Experimental Setup}

\noindent We use the convolutional layers of ResNet-50 pretrained on ImageNet \cite{resnet2016} to perform the training. The models are optimized using Adam \cite{adam2014}, we set the initial learning rate $10^{-5}$, weight decay 4 × $10^{-4}$. During training, we randomly sample \textit{k} classes and $|P|$ samples per class to form each mini-batch. Following standard practice, we resize images to 256 × 256, and randomly crop them to 224 × 224 as input. Random flipping ($p = 0.5$) is used during training for data augmentation, and a single center crop of 224 × 224 is used during evaluation. We use 0 as the fixed random seed for all experiments to avoid seed-based performance fluctuations.  During training, we directly set $\tau$ in Eq.~\ref{g_function} to 0.01 which is explored in \cite{2020smoothap}.

For all the datasets, every instance from each class is used in turn as the query $I_q$, and the retrieval set $\Omega$ is formed as all the remaining instances. Recall@k (R@k) is adopted  as the main evaluation metric. In order to evaluate the generalization of the model, we also display dists@intra (Mean Intraclass Distance), dists@inter (Mean Interclass Distance) \cite{roth2020revisiting} and Normalized Mutual Information (NMI) \cite{schutze2008introduction} for further performance analysis.  BS represents mini-batch size.

\subsection{Performance of Different PNP Variants}

\noindent In this section, we investigate different variants of PNP on two popular benchmark datasets. The batch size of experiments in Table~\ref{form_PNP} and Table~\ref{form_vehicle} are both 112. We report the best performance of different variants and the impact of hyper-parameters is shown in Fig.~\ref{parameters}. 

We present R@k, Distance and NMI in Table~\ref{form_PNP} and Table~\ref{form_vehicle}. PNP-D consistenly outperforms PNP-O and PNP-I on two benchmark datasets for R@k and achieves larger dists@intra, dists@inter and NMI, which shows better generalization of the model \cite{roth2020revisiting}. Specifically, PNP-D is 2.1\% higher than PNP-O on SOP benchmark and 0.9\% higher on VehicleID for R@1. In contrast, PNP-I is worse than PNP on R@1, and it is about 1\% lower than PNP on SOP benchmark 1-2\% on VehicleID for R@1. 

\begin{figure}[t]

\centering

\includegraphics[width=7.5cm]{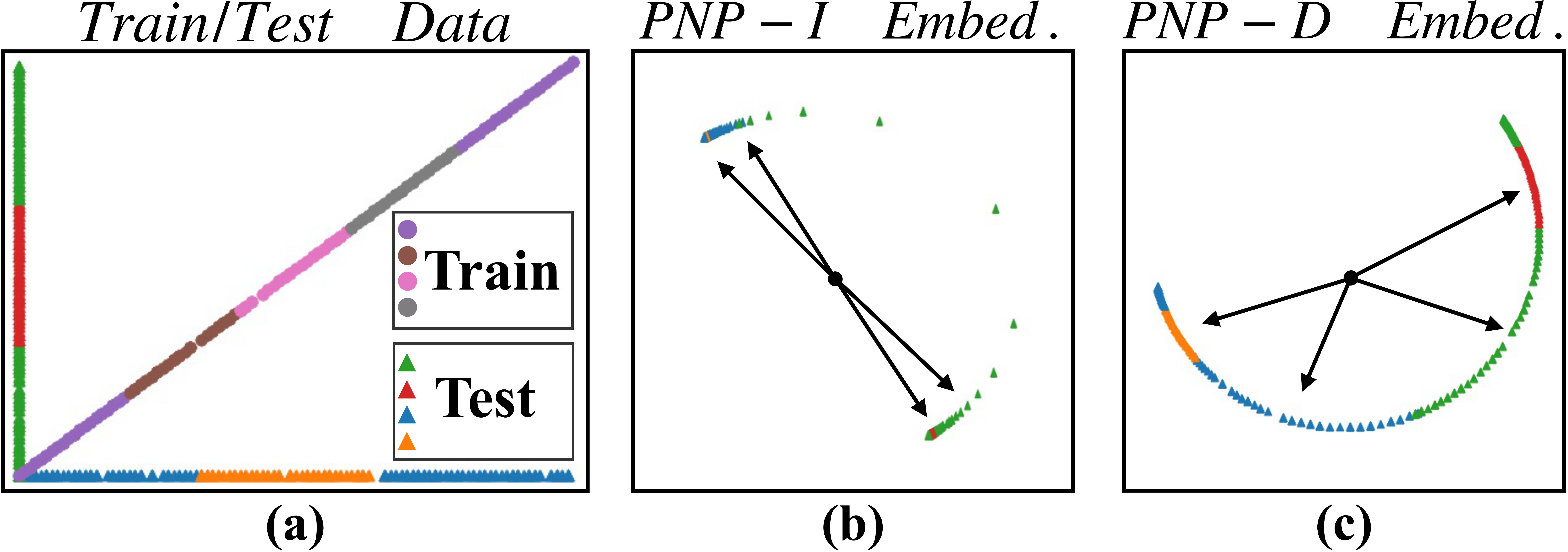}

\caption {Illustration of the effect of different variants. (a) training and test data (b) The network trained with PNP-I fails to separate all test classes due to the aggregation of all relevant instances. (c) The PNP-D successfully separates the test classes by keeping the intra-class variance.}
\label{toy_distribute}

\end{figure}

PNP-I assigns larger gradients to positive instances which have more negative ones before, thus it will enforce relevant images to gather and result in smaller dists@intra and dists@inter. In contrast, PNP-D assigns smaller gradients to such instances and slowly optimizes them by considering that they have a high probability of belonging to other centers of the corresponding category, and therefore it has larger dists@intra and dists@inter. By this strategy, PNP-D can adaptively achieve “multi-centers” for each category if necessary and automatically adjust the number of clusters. Such strategy also solves the error correction of noise by considering the noisy data as another center.

Previous works have found that larger dists@intra and dists@inter will produce better generalization \cite{roth2020revisiting}.  The results shown in Table~\ref{form_PNP} and Table~\ref{form_vehicle} further confirm such conclusion. Compared to PNP-I, PNP-D has larger dists@intra and dists@inter, and results in larger NMI, which shows better generalization of the model. Similar to \cite{roth2020revisiting}, we conduct one experiment to illustrate the performance of two loss functions.  Specifically, we use a fully-connected network with two layers and 30 neurons for each layer. The dimensions of input and embeddings are both 2D, and the embeddings are normalized on a unit circle. As shown in  Fig.~\ref{toy_distribute} (a), each of four training and test classes contains 150 samples, respectively. We train the networks using PNP-I$_b$ and PNP-D$_q$, respectively. Fig.~\ref{toy_distribute} illustrates the effect of two loss functions. PNP-I fails to separate all test classes in Fig.~\ref{toy_distribute} (b) while PNP-D successfully separates them (with larger intra-class variance) in Fig.~\ref{toy_distribute} (c) . It shows that PNP-D enables the model to capture more information and exhibits stronger generalization to the unseen test classes by retaining the variance within the class in the training process.

\begin{figure}[t]

\centering

\includegraphics[width=7cm]{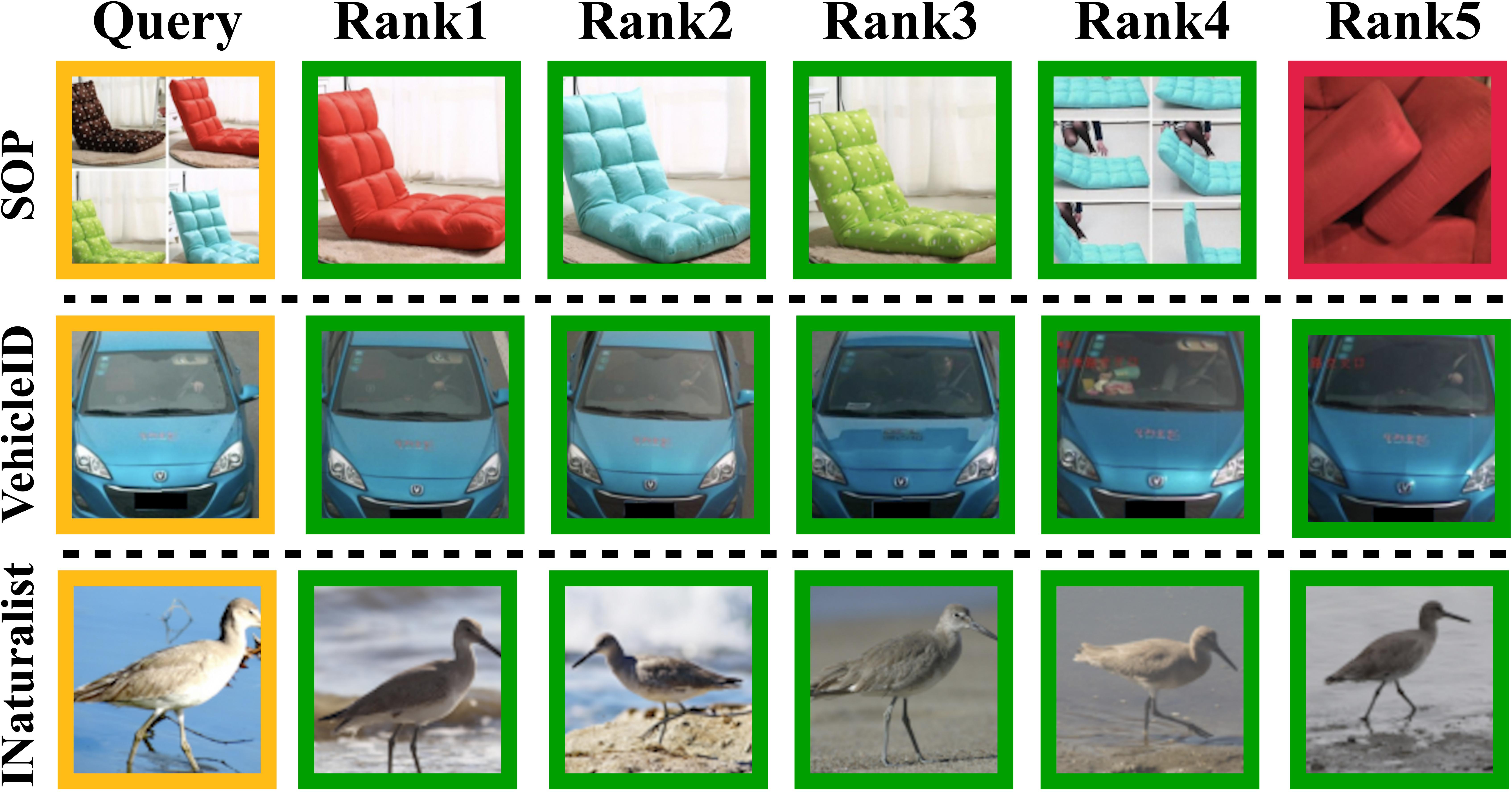}

\caption{Image retrieval examples on three datasets. Images with green border are positive instances and these with red border are negative ones for the query.}
\label{rank_example}

\end{figure}

\newcommand{\tabincell}[2]{\begin{tabular}{@{}#1@{}}#2\end{tabular}}
\begin{table}[t]

\liuhao

\centering


\setlength{\tabcolsep}{0.6mm}{
\begin{tabular}{c|cccc}

\bottomrule

Method & R@1 & R@10 & R@100 & R@1K \\ \hline

\multicolumn{1}{l|}{Hist. \cite{2016histogramloss}} & 72.4& 86.1 &94.1 &98.3\\ 

\multicolumn{1}{l|}{Margin \cite{2017margin}} & 72.7& 86.2 &93.8 &98.0\\ 

\multicolumn{1}{l|}{Divide \cite{2019divide}} & 75.9&88.4&94.9&98.1\\ 

\multicolumn{1}{l|}{FastAP \cite{2019fastap}} &76.4 &89.0 &95.1 &98.2\\ 

\multicolumn{1}{l|}{MIC \cite{2019tripleth}} &77.2 &89.4 &95.6 &-\\ 

\multicolumn{1}{l|}{SoftTriplet \cite{2019softtriple}} &78.3 &90.3 &95.9 &-\\ 

\multicolumn{1}{l|}{RankMI \cite{2020RankMI}} &74.3 &87.9 &94.9 &98.3\\ 

\multicolumn{1}{l|}{Blackbox AP \cite{2020blackboxap}} &78.6 &90.5 &96.0 &98.7\\ 

\multicolumn{1}{l|}{Cont. w/M \cite{2020cont}} & 80.6 &91.6 &96.2 &98.7\\ 


\multicolumn{1}{l|}{Pnca++ \cite{2020proxynca++}} &80.7 &92.0 &96.7 &98.9 \\

\multicolumn{1}{l|}{Smooth-AP \cite{2020smoothap}} &80.1 &91.5 &96.6 &99.0 \\ 

\multicolumn{1}{l|}{DCML-MDW \cite{Zheng_2021_CVPR}} &79.8 &90.8 &95.8 &- \\ \hline
\multicolumn{1}{l|}{PNP-D$_q$ (BS = 112)} & 80.1 & 91.5 & 96.7 & 99.0 \\ 


\multicolumn{1}{l|}{PNP-D$_q$ (BS = 384)} &\textbf{81.1} &\textbf{92.2} &\textbf{96.8} &\textbf{99.0} \\ \bottomrule

\end{tabular}}
\caption{Performance comparison  on {SOP} (\%).}
\label{sota_sop}

\end{table}

\subsection{Comparison with State-of-the-art}

\noindent We compare PNP loss to the recent AP-based methods and a series of state-of-the-art deep metric learning methods on three standard benchmarks. 
\subsubsection*{SOP}

\indent\setlength{\parindent}{1em}For fair comparison, we use the same setting with \cite{2020smoothap}. Table~\ref{sota_sop} shows the performance on SOP, we observe that PNP-D$_q$ achieves the state-of-the-art results. In particular, our model outperforms other methods on all evaluation metrics and outperforms AP approximating methods  by 0.5 - 4\% on R@1 when the same batch size (384) and dimension (512) are used. Note that, although current work (Cont. w/M \cite{2020cont}) uses memory techniques to sample from many mini-batches simultaneously for each iteration, our best model still outperforms this method, only using a single mini-batch on each iteration. 

\begin{table}[t]

\liuhao

\centering


\setlength{\tabcolsep}{0.8mm}{
\begin{tabular}{c|cc|cc|cc}

\bottomrule

\multirow{2}{*}{\moren{Method}}&\multicolumn{2}{c|}{Small} & \multicolumn{2}{c|}{Medium} &\multicolumn{2}{c}{Large}\\ \cline{2-7}
& R@1& R@5 &R@1& R@5 &R@1& R@5 \\ \hline

\multicolumn{1}{l|}{Divide \cite{2019divide}} & 87.7 &92.9 &85.7& 90.4& 82.9 &90.2\\ 

\multicolumn{1}{l|}{MIC. \cite{2019tripleth}} & 86.9 &93.4&-&-& 82.0 &91.0\\ 

\multicolumn{1}{l|}{FastAP \cite{2019fastap}} &91.9& 96.8& 90.6& 95.9& 87.5& 95.1\\ 

\multicolumn{1}{l|}{Cont. w/M \cite{2020cont}} & 94.7& 96.8& 93.7& 95.8& 93.0& 95.8\\ 

\multicolumn{1}{l|}{Smooth-AP \cite{2020smoothap}} &94.9 & 97.6 & 93.3 & 96.4 & 91.9 & 96.2 \\ \hline

\multicolumn{1}{l|}{PNP-D$_q$ (BS = 112)}&94.6&97.4&92.9&96.3&91.4&95.9 \\

\multicolumn{1}{l|}{PNP-D$_q$ (BS = 384)}& \textbf{95.5} & \textbf{97.8} &\textbf{94.2}& \textbf{96.9} & \textbf{93.2} & \textbf{96.6}  \\ \bottomrule

\end{tabular}}
\caption{Performance comparison  on {VehicleID} (\%).}
\label{sota_VID}

\end{table}

\begin{table}[t]

\liuhao

\centering


\setlength{\tabcolsep}{1.8mm}{

\begin{tabular}{c|cccc}

\bottomrule

Method & R@1 & R@4 & R@16 & R@32 \\ \hline

\multicolumn{1}{l|}{Triplet \cite{2017margin}} & 58.1& 75.5 &86.8 &90.7\\ 

\multicolumn{1}{l|}{Proxy NCA \cite{2017pcna}} &61.6&77.4&87.0&90.6\\ 

\multicolumn{1}{l|}{FastAP \cite{2019fastap}} &60.6 &77.0 &87.2 &90.6\\ 

\multicolumn{1}{l|}{Blackbox AP \cite{2020blackboxap}} &62.9 &79.0 &88.9&92.1\\ 

\multicolumn{1}{l|}{Smooth-AP \cite{2020smoothap}} & 65.9 & 80.9 & \textbf{89.8} & \textbf{92.7} \\ \hline

\multicolumn{1}{l|}{PNP-D$_q$ (BS = 224)} &\textbf{66.6} &\textbf{81.1} &89.7 & 92.6 \\ \bottomrule


\end{tabular}}
\caption{Performance comparison  on {INaturalist} (\%).}
\label{sota_INaturalist}

\end{table}

\begin{table}[t]

\liuhao

\centering


\setlength{\tabcolsep}{4mm}{
\begin{tabular}{c|cccc}

\bottomrule

Method & R@1 & R@10 & R@100 & R@1K \\ \hline

\multicolumn{1}{l|}{Smooth-AP} &73.2&86.4&93.6&97.5\\ 

\multicolumn{1}{l|}{PNP-I$_u$} &67.5&82.7&92.3&97.5\\ 

\multicolumn{1}{l|}{PNP-D$_q$} &\textbf{73.8}&\textbf{87.1}&\textbf{94.2}&\textbf{98.0} \\ \bottomrule

\end{tabular}}
\caption{Evaluation of robustness on {SOP} (\%).}
\label{robustness_sop}

\end{table}

\begin{table}[t]

\liuhao

\centering


\setlength{\tabcolsep}{2.3mm}{
\begin{tabular}{c|cc|cc|cc}

\bottomrule

\multirow{2}{*}{\moren{Method}}&\multicolumn{2}{c|}{Small} & \multicolumn{2}{c|}{Medium} &\multicolumn{2}{c}{Large}\\ \cline{2-7}
& R@1& R@5 &R@1& R@5 &R@1& R@5 \\ \hline

\multicolumn{1}{l|}{Smooth-AP} &89.9&95.8&88.6&94.3&85.2&93.2\\ 

\multicolumn{1}{l|}{PNP-I$_u$}  &86.9&94.0&85.0&92.6&80.8&90.7\\ 

\multicolumn{1}{l|}{PNP-D$_q$} &\textbf{92.6}&\textbf{96.3}&\textbf{90.8}&\textbf{95.3}&\textbf{88.6}&\textbf{94.6}  \\ \bottomrule

\end{tabular}}
\caption{Evaluation of robustness on {VehicleID} (\%).}
\label{robustness_VID}

\end{table}

\begin{figure}[t]
\centering
\includegraphics[width=8.3cm]{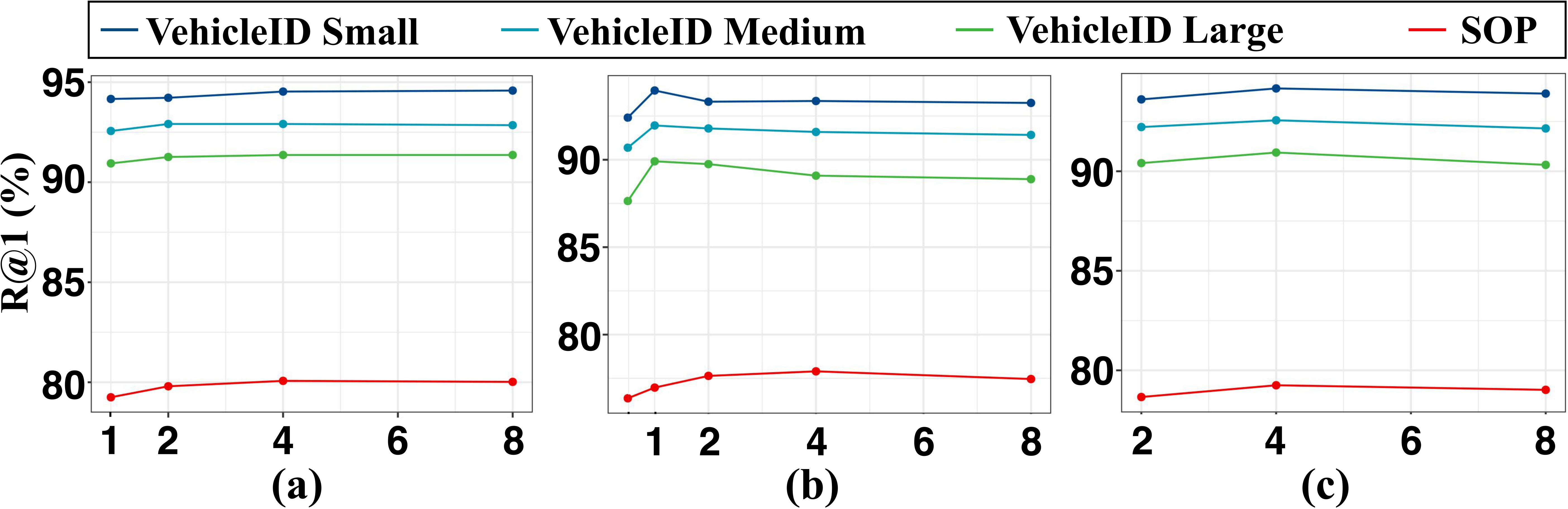}
\caption{Impact of different hyper-parameters on SOP and VehicleID. (a) steepness $\alpha$, (b) boundary b, (c) samples per class during mini batch sampling $|P|$. }
\label{parameters}
\end{figure}

	

\subsubsection*{VehicleID}

\indent\setlength{\parindent}{1em}We further conduct experiments on VehicleID to verify the performance of PNP on large-scale retrieval datasets. Table~\ref{sota_VID} shows the results on the VehicleID dataset. For the same batch size (384) and dimension (512), we observe that PNP-D$_q$ again achieves state-of-the-art performance on the large-scale VehicleID dataset on all the evaluation metrics. Such result shows that PNP-D also works better on large-scale datasets. 

\subsubsection*{INaturalist}

As shown in Table~\ref{sota_INaturalist}, our method also outperforms the best method by 0.7\% on R@1 for the experiments on INaturalist with the same batch size (224) and dimension (512). These results demonstrate that PNP is particularly suitable for large-scale retrieval datasets, which demonstrates its scalability to real-world retrieval problems.

\subsection{Impact of Hyper-parameters}

\noindent To investigate the effect of different hyper-parameter settings, \textit{i.e.}, steepness $\alpha$ in Eq.~\ref{reci}, boundary $b$  in Eq.~\ref{b}, samples per class $|P|$, we train ResNet-50 on SOP and VehicleID. The random seed is fixed to 0 and the batch size is set to 112 for all experiments in this section.

\noindent \textbf{Steepness $\alpha$}: The influence of $\alpha$ of PNP-D$_q$ is shown in Fig.~\ref{parameters} (a). We can find that larger $\alpha$ results in larger R@1. It can be explained that larger $\alpha$ corresponds to more stable optimization (correct positive instances with sufficient confidence), and thus achieves better performance. However, when $\alpha$ is too large, the training phrase will crash due to the gradient exploding. Therefore, we use as larger $\alpha$ as possible to get better performance.

\noindent \textbf{Boundary} $b$: The influence of $b$ for PNP-I$_b$ is shown in Fig.~\ref{parameters} (b). $b = 4$ is the best on SOP while $b = 1$ is the best on VehicleID. $b$ controls the biggest gradient of the PNP-I$_b$ with respect to $R(i,S_N)$, and different datasets require different adjusting range. We report the best value for the comparison of different variants.

\noindent \textbf{Samples Per Class $|P|$}: The influence of the samples per class $|P|$ in a mini-batch is shown in Fig.~\ref{parameters} (c). We observe that $|P|$ = 4 results in the highest R@1. The probable reason is that mini-batches are formed by sampling from each class, where a low value means a larger number of sampled classes and a higher probability of sampling hard-negative instances that violate the correct ranking order. Increasing the number of classes in the batch results in a better batch approximation of the true class distribution, allowing each training iteration to enforce a more optimally structured embedding space. However, too small value ($|P|$ = 2) leads to worse performance because it can not give enough positive instances to violate the correct ranking order for training.

\subsection{Evaluation of Robustness}

\noindent In order to evaluate the robustness and the performance of multi-centers per class, we perform extensive experiments on constructed datasets in this section. Specifically, we use the train set of SOP and VehicleID as benchmark datasets and merge every three categories to form one new class with multi-centers (at least three centers per class). As for evaluation, we still use original test set to evaluate the model. During the training, we use PNP-I$_b$, PNP-D$_q$ and Smooth-AP as loss functions.

The results are shown in Table~\ref{robustness_sop}, \ref{robustness_VID}. Compared with using original training set, these three methods all have performance loss on two benchmark datasets. Generally, PNP-D$_q$ and Smooth-AP with similar gradient assignment strategy have less performance loss than PNP-I$_b$ because of their adaptability to multi-centers, and PNP-D achieves the best performance. Specifically, for VehicleID, PNP-D exceeds the performance of Smooth-AP by 2 - 3\% and PNP-I by a significant 6 - 8\% on R@1.

The above results show better robustness of PNP-D. Noisy data is inevitable in the real-world data and it will leads to the instability of training. The above experiments show that PNP-D can achieve comparable performance, even if there are a lot of wrong annotations in datasets. This is because PNP-D can automatically remove the interference of noisy data by putting it in a single center.

\section{Conclusions}
\noindent In this paper, we propose a novel PNP loss, which directly improves the retrieval performance by penalizing negative instances before positive ones. Moreover, we find that different derivative functions of losses correspond to different gradient assignments. Therefore, we systematically investigate different gradient assignment solutions via constructing derivative functions of losses, resulting in PNP-I and PNP-D. PNP-D consistently achieve state-of-the-art performance on three benchmark datasets. We just provide four variants and will explore more forms via similar strategy in the future. In addition, we hope the proposed strategy for designing loss functions can also be applied to other tasks.

\section{Acknowledgments}
\noindent This work was supported in part by Beijing Natural Science Foundation under Grant L182054, in part by the National Natural Science Foundation of China under Grant 62125207, 61972378, 61532018, U1936203, U19B2040. This research was also supported by Meituan-Dianping Group.

{
\bibliography{aaai22}
}

\end{document}